\documentclass[conference]{IEEEtran}
\IEEEoverridecommandlockouts
\usepackage{cite}
\usepackage{amsmath,amssymb,amsfonts}
\usepackage{algorithmic}
\usepackage{graphicx}
\usepackage{textcomp}
\usepackage[format=plain, labelfont={bf,it}]{caption}
\usepackage{xcolor}
\usepackage{authblk}
\usepackage{listings}
\usepackage{makecell}
\usepackage{float}
\usepackage{csvsimple}
\usepackage{footnote}
\usepackage{numprint}
\usepackage{llvm/lang}  
\usepackage{nasm/style} 
\newfloat{lstfloat}{htbp}{lop}
\floatname{lstfloat}{Listing}
\usepackage{xcolor}
\usepackage{tikz}

\def\BibTeX{{\rm B\kern-.05em{\sc i\kern-.025em b}\kern-.08em
    T\kern-.1667em\lower.7ex\hbox{E}\kern-.125emX}}

\newcommand\copyrighttext{%
  \footnotesize \textcopyright 2023 IEEE. Personal use of this material is permitted. Permission from IEEE must be obtained for all other uses, in any current or future media, including reprinting/republishing this material for advertising or promotional purposes, creating new collective works, for resale or redistribution to servers or lists, or reuse of any copyrighted component of this work in other works.}
\newcommand\copyrightnotice{%
\begin{tikzpicture}[remember picture,overlay]
\node[anchor=south,yshift=10pt] at (current page.south) {\fbox{\parbox{\dimexpr\textwidth-\fboxsep-\fboxrule\relax}{\copyrighttext}}};
\end{tikzpicture}%
}
    
\begin{document}

\title{Fortran High-Level Synthesis: Reducing the barriers to accelerating HPC codes on FPGAs}

\author[1]{Gabriel Rodriguez-Canal \textsuperscript{\textsection}}
\author[1]{Nick Brown}
\author[2]{Tim Dykes}
\author[2]{Jessica R. Jones}
\author[2]{Utz-Uwe Haus}
\affil[1]{EPCC, The University of Edinburgh}
\affil[2]{HPC/AI EMEA Research Lab, Hewlett Packard Enterprise}
\maketitle
\copyrightnotice

\begin{abstract}
In recent years the use of FPGAs to accelerate scientific applications has grown, with numerous applications demonstrating the benefit of FPGAs for high performance workloads. However, whilst High Level Synthesis (HLS) has significantly lowered the barrier to entry in programming FPGAs by enabling programmers to use C++, a major challenge is that most often these codes are not originally written in C++. Instead, Fortran is the lingua franca of scientific computing and-so it requires a complex and time consuming initial step to convert into C++ even before considering the FPGA.

In this paper we describe work enabling Fortran for AMD Xilinx FPGAs by connecting the LLVM Flang front end to AMD Xilinx's LLVM back end. This enables programmers to use Fortran as a first-class language for programming FPGAs, and as we demonstrate enjoy all the tuning and optimisation opportunities that HLS C++ provides. Furthermore, we demonstrate that certain language features of Fortran make it especially beneficial for programming FPGAs compared to C++. The result of this work is a lowering of the barrier to entry in using FPGAs for scientific computing, enabling programmers to leverage their existing codebase and language of choice on the FPGA directly.
\end{abstract}

\begingroup\renewcommand\thefootnote{\textsection}
\footnotetext{Corresponding author: gabriel.rodcanal@ed.ac.uk}

\section{Introduction}
First introduced in the late 1950s, Fortran has been the lingua franca of scientific computing for over 60 years \cite{brainerd2003importance}. Whilst many alternative programming languages have come and gone, it has regained its popularity for writing high performance codes. Indeed, over 80\% of the applications running on ARCHER2, a 750,000 core Cray EX which is the UK national supercomputer, are written in Fortran. This ubiquity has meant that there are many applications, both actively maintained and legacy, that use the language ranging from weather forecasting models used by the world's leading meteorology organisations \cite{strazdins2011profiling} \cite{brown2020highly} to simulation codes used for modelling state of the art aircraft and jet engines \cite{yang2019applicability} \cite{lehmkuhl2020active}. 

As the HPC community turns to exploring accelerators other than GPUs, for potential performance and energy usage advantages, Field Programmable Gate Arrays (FPGAs) have significant potential. Whilst FPGAs are yet to gain wide acceptance in HPC, there are a number of testbeds and smaller HPC systems that contain FPGAs. Numerous studies report that the ability to tailor the logic and select a bespoke on-chip memory configuration can be beneficial compared to general purpose architectures, especially for those codes which are memory bound \cite{brown2021accelerating} \cite{davila2020analytical}. However, a major blocker is that currently, when using High Level Synthesis (HLS), one must write their codes in C or C++. Consequently there is a significant initial overhead for many HPC developers where they must first convert their Fortran code into C++, before getting anywhere near an FPGA. This not only requires expertise in another language, but furthermore is a time consuming and error prone process due to the differences between Fortran and C++, for instance in array index ordering (row major in C verses column major in Fortran), and default array start indices.

This paper describes work bringing Fortran programming to FPGAs by integrating Flang with AMD Xilinx's LLVM-based HLS back end. The objective is to lower the barrier to entry in porting existing HPC codes to FPGAs, and ultimately increase the accessibility of FPGAs to the HPC community. The rest of the paper is structured as follows, Section \ref{sect:background} describes the background to this work in more detail and surveys previous efforts bringing Fortran to FPGAs. Section \ref{sec:experiment} reports the hardware and software configurations used in this work, before Section \ref{sect:implementation} then explores our Fortran HLS solution. A performance comparison of our work against the existing C/C++ HLS approach is undertaken in Section \ref{sect:evaluation}, highlighting benefits of Fortran for writing HLS code. Section \ref{sec:conclusions} then draws conclusions and describes further work.

The contributions of this paper are:
\begin{enumerate}
    \item Describing an approach to enabling direct support for Fortran on AMD Xilinx FPGAs through the Xilinx Vitis ecosystem.
    \item A case study of leveraging AMD Xilinx's open source LLVM front end components to integrate new programming languages and tools into Vitis HLS, without changes needed to the standard Flang front end or Xilinx back end.
    \item A comparison of performance between Fortran HLS and C/C++, demonstrating how Fortran specific language constructs, such as dynamic N-dimensional arrays, can provide an advantage over the existing HLS C/C++ flow.    
\end{enumerate}


\section{Background and related work} \label{sect:background}
A common approach to programming FPGAs is to use C or C++ via HLS, with both AMD Xilinx and Intel providing their own HLS solutions. Several tools have been developed \cite{ye2022scalehls} \cite{curzel2022higher} which aim to undertake automatic code transformations, often via bespoke MLIR dialects \cite{lattner2021mlir}, in order to enable non-experts to program FPGAs effectively. There are mixed results when using these tools, and code must be in C or C++ to start with, but it demonstrates the ability of third parties to develop their own transformations and then integrate with existing tools.


DACE \cite{ziogas2021productivity} is an alternative approach which is more focused on HPC, where developers write code in Python and this is then transpiled to a variety of target architectures including FPGAs (C/C++ HLS for AMD Xilinx and OpenCL for Intel FPGAs). However, the limitation is that one must first undertake conversion of their code to integrate with DACE before running on FPGAs, similar to the disadvantages associated with having to first convert Fortran to C/C++ for HLS. 

TyTra \cite{nabi2019automatic} is a compiler which synthesises Fortran to HDL and is intended as a platform for undertaking auto-optimisation steps to automatically convert the programmer's Von Neumann-based code into an efficient dataflow representation. Whilst, at the time of writing TyTra is not actively maintained, it is an interesting comparison point. The major disadvantages with TyTra is that one must rely on their own bespoke compiler resulting in risk (whether the compiler will continue to be maintained in the future), potentially poorer future performance (as they will not automatically benefit from the improvements made by vendors to their tool chains) and a lack of integration with the wider ecosystem (such as the Vitis profiling tools).

\subsection{AMD Xilinx's LLVM-based HLS tooling}

AMD Xilinx's  HLS synthesis tool is based upon LLVM and, as illustrated in Figure \ref{fig:default_workflow}, this comprises two parts; a front end built upon Clang \cite{lattner2008llvm} which compiles a programmer's C/C++ code to LLVM's Intermediate Representation format known as LLVM-IR, and a back end which synthesises LLVM-IR to the underlying HDL level for their FPGAs. Whilst much of Clang in Vitis HLS has remained unchanged, AMD Xilinx have added support for their own bespoke HLS pramas which correspond to additional constructs in the generated LLVM-IR. 

\begin{figure}
\centering
\includegraphics[width=\linewidth]{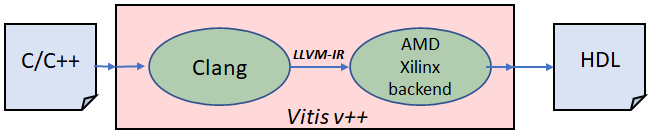}
\caption{Default workflow of AMD Xilinx Vitis \texttt{v++}}	
\label{fig:default_workflow}
\end{figure}

\begin{figure*}
\centering
\includegraphics[width=\linewidth]{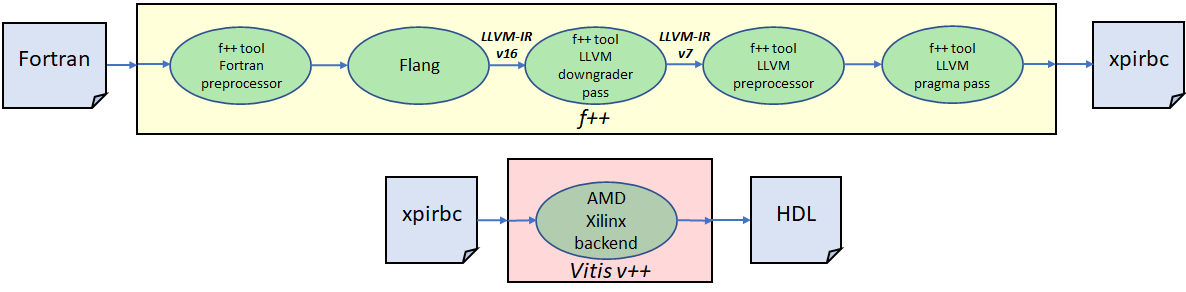}
\caption{Overall workflow of our Fortran, \texttt{f++}, tool where Fortran is converted into the \emph{xpirbc} format by our tool and this is then provided to the AMD Xilinx back end which generates the HDL.}	
\label{fig:fpp_workflow}
\end{figure*}

AMD Xilinx have released their HLS Clang front end via the Illinois/NCSA open source licence, and this is kept up to date with each release of the Vitis suite, although their back end remains proprietary. Based upon these components it is possible to generate LLVM-IR and provide this to the AMD Xilinx back end for it to synthesise to HDL.

\subsection{LLVM Flang}

Flang \cite{kedward2022state} is the open source Fortran compiler provided as part of the LLVM compiler suite. Based around an active and growing open source community, this compiler was developed from the ground-up around five years ago and replaced the previous Flang compiler in the main LLVM repository around two years ago. Built around MLIR, a framework in LLVM for defining bespoke IR dialects and mixing these together, Flang currently supports the Fortran 2003 standard with several features from later standards, such as Fortran 2008 and Fortran 2018, also supported and many others in development.

Ultimately, Flang is a front end for the Fortran language and after language specific lexing, parsing, and optimisation, generates LLVM-IR. This is then passed to a specific LLVM back end and built into the final executable, for instance targeting CPUs or GPUs.

\section{Experimental setup}
\label{sec:experiment}
An AMD Xilinx U280 FPGA is used for the FPGA experiments conducted in this paper, at the default clock frequency of 300MHz. This contains an FPGA chip with 1.08 million LUTs, 4.5MB of on-chip BRAM, 30MB of on-chip UltraRAM, and 9024 DSP slices. This PCIe card also contains 8GB of High Bandwidth Memory (HBM2) and 32GB of DDR DRAM. All bitstreams are built for the U280 using Xilinx's Vitis framework version 2021.2 which at the time of writing is the latest compatible version for the U280. We use Flang version 16 throughout. Reported runtime is the execution time of the HLS kernel on the FPGA and excludes the overhead of data transfer to or from the device. Kernel execution time is gathered via profiling information on the OpenCL event.

For the CPU comparison runs, which are intended to provide context around the optimisations performed by the Clang and Flang front ends respectively, were undertaken on a single core of a Xeon Platinum (Cascade Lake) 8260M CPUs running at 2.40GHz. All reported performance times are based on wall clock time and numbers are averaged over 10 runs.

\section{Fortran FPGA programming via Flang HLS} 
\label{sect:implementation}

The behaviour of AMD Xilinx's Vitis \texttt{v++}, the tool that synthesises C/C++ code by calling HLS, can be altered to provide AMD Xilinx's LLVM-based back end externally generated LLVM-IR as long as the file has the extension \texttt{.xpirbc}. We leverage this approach to enable Fortran on Xilinx's FPGAs by connecting LLVM Flang to the AMD Xilinx back end via Flang's generated LLVM-IR, with the major objective being to leave LLVM Flang and the AMD Xilinx back end unchanged.

There are numerous challenges that must be overcome for AMD Xilinx's HLS back end to properly interpret the LLVM-IR. This is because Flang must conform to certain requirements when generating LLVM-IR based upon what is expected by the AMD Xilinx back end. In summary, these are:
\begin{enumerate}
    \item The generated LLVM-IR must conform to LLVM version 7. 
    \item HLS pragmas must be lowered into AMD Xilinx's specific LLVM-IR metadata or directives.
    \item HLS Streams are defined and managed with custom IR primitives generated by AMD Xilinx's custom Clang front end.
\end{enumerate}

The challenge is that none of these requirements are directly met by Flang. For instance, at the time of writing the latest version of LLVM is version 16, in fact version 7 was from 2018 and is not supported by the latest Flang. Similarly, Flang has no internal knowledge of HLS streams or HLS annotations and as a result, the IR that is generated is not compatible with the back end unless these are explicitly added. 

To address these issues, our approach places a layer between Flang and AMD Xilinx's back end which, as illustrated in Figure \ref{fig:fpp_workflow}, performs passes and regular expressions on the Flang generated LLVM-IR to transform it into a form that is acceptable to the back end. These are all wrapped in our \texttt{f++} tool, which launches separate components of the Fortran HLS workflow.

\subsection{Lowering the generated LLVM-IR to version 7}

The differences between the latest versions of LLVM-IR and version 7 will either result in the LLVM-IR generated by Flang being rejected by the AMD Xilinx HLS back end, or accepted and subsequent erroneous results. The later can be especially challenging, for instance if the LLVM-IR contains the \texttt{byval} attribute for memory pointers, which is used by the latest LLVM version, then HLS will succeed and generate HDL, but after running the synthesis, place, and route stages the resulting  bitstream will be incompatible with the Xilinx Runtime library, XRT. Whilst it might seen an obvious approach for Xilinx to update their LLVM back end to accept the latest LLVM-IR, this is non-trivial because of the differences between LLVM-IR version 7 and the current version. An example of this is that, as of version 16, LLVM-IR only supports opaque pointers at the IR level which are incompatible with HLS. Furthermore, the rapid pace of change in and around LLVM means that it is not always possible for mature products, such as AMD Xilinx's HLS, built upon this technology to track the latest bleeding-edge version.

Therefore, we developed a pass that will downgrade the LLVM-IR to version 7. This was initially based upon the prototype developed by the Khronos team \cite{forget2022single} as part of their integration with SYCL. However, their work was built for Clang and only partially amends the generated LLVM-IR. For example, our pass also transforms procedure names into the format expected by the back end.Furthermore, to enable compatibility with Flang our pass also transforms procedure argument names, which cannot begin with a number, but are generated as such by Flang. Finally, we strip the LLVM-IR of all metadata which does not conform to LLVM-IR version 7. 

\subsection{HLS pragmas}
Our research prototype supports a subset of the most commonly used HLS pragmas, and it is trivial to add others following the same approach we have developed. It was decided to concentrate on those pragmas which are fundamental for controlling performance, namely the \emph{pipeline}, \emph{unroll}, \emph{array\_partition}, and \emph{dataflow} pragmas, as well as the \emph{interface} pragma which enables programmers to define the data interfaces of the generated HLS IP block. Crucially, all pragmas are handled in a similar way, and supporting a pragma involves the following steps:

\begin{enumerate}
    \item Preprocessing to replace the pragma by a call to a placeholder Fortran subroutine with no arguments. This will lead to the Flang compiler generating a placeholder function call in the LLVM-IR, and we followed this approach because it does not introduce extra instructions before or after the call. Any arguments passed to the pragma are encoded as part of the subroutine name and this approach has been adopted because, if we were to pass them as arguments, then Flang will generate corresponding values in the IR, which makes it more difficult to then process and can in some circumstances result in bottlenecks in the generated HDL.
    \item Run our bespoke LLVM pass which operates on the Flang generated LLVM-IR. Each placeholder subroutine call is identified and any arguments are decoded and extracted. The corresponding LLVM-IR constructs for the specific pragma that this corresponds to are inserted into the IR as appropriate.
    \item Remove placeholder subroutine calls from the LLVM-IR    
\end{enumerate}

We followed this approach because it means that we can undertake pragma replacement via regular expressions in a preprocessing phase, rather than having to modify Flang itself to be aware of the pragmas. This would have required changes to Flang which would then diverge it from the main tree.

An important aspect to highlight, and a further challenge, is how the contention of instructions operate in loop trees in LLVM-IR, as this is needed to correctly correlate the \emph{pipeline} and \emph{unroll} pragmas with the loops that they are operating over. In LLVM, given a loop $L$ and any of its sub loops $S$ and the instruction $I$, $I \in S \implies I \in L$. Therefore, our pass must identify the deepest sub loop in the loop tree that contains $I$ using depth first search with backtracking. Otherwise, for nested loops, if the corresponding LLVM-IR was generated at the location of the place holder subroutine call, then the parent loop would be qualified in every case instead, altering the semantics of the program.  Figure \ref{fig:loop_tree} illustrates the equivalence of a series of nested \emph{for} loops and their corresponding loop tree, where we undertake this search for the annotated loop.

\begin{figure}
\centering
 \includegraphics[width=0.5\textwidth]{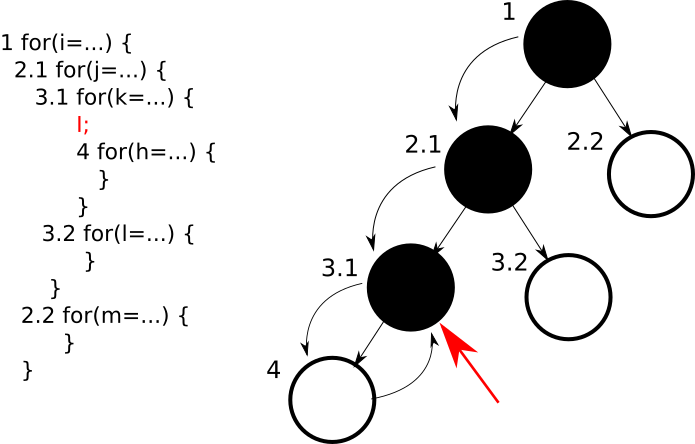}
\caption{Nested for loops and their loop tree, traversed with DFS with backtracking. Pointed by the red arrow, the loop containing instruction I. Black nodes contain the instruction.}	
\label{fig:loop_tree}
\end{figure}

\subsection{HLS streams and polymorphism}

In AMD Xilinx's HLS, the C++ template mechanism is leveraged to implement the HLS streams library and this uses polymorphism to enable many different datatypes to be streamed. Whilst polymorphism is available from Fortran 2008 onwards, we decided to keep our tool backwards-compatible with Fortran 90 because that is the most common Fortran version used by HPC codes. Consequently, it was undesirable to encode any features that required later versions of Fortran. Therefore, we replicate the polymorphic behaviour with a library of preprocessor macros and preprocessing of the Fortran code using regular expressions. 

\begin{lstfloat}
\begin{lstlisting}[language=Fortran, frame=lines, label=lst:code_example, numbers=left, caption=Example Fortran program using our Fortran HLS Stream macro library.]
proto_hls_stream(integer)

subroutine f(a, b)
    use hls_stream

    integer, dimension(100) :: a, b
    type(HLSStream) :: s

    set_hls_stream_type(s, integer)

    do i = 1,100
        call hls_write(a(i), s)
    end do

    do i = 1,100
        b(i) = hls_read(s)
    end do
end subroutine
\end{lstlisting}
\end{lstfloat}

Listing \ref{lst:code_example} sketches an example of Fortran code using our macro-based HLS Stream library. The stream functions provided in Vitis HLS (\texttt{read}, \texttt{write}, \texttt{empty} and \texttt{full}) are also provided in Fortran HLS. The module is instantiated in line 1 for the types that will be used with streams in the code, just an \texttt{integer} in this example, with twenty Fortran types and precisions supported overall by the macro. Listing \ref{lst:module_instantiation} illustrates the macro expansion which will replace line 1 in Listing \ref{lst:code_example} by this preprocessor generated code. 

The generated \emph{hls\_stream} module contains the user derived type \texttt{HLSStream} which enables Flang to check the correct use of the library at compile-time. For the types provided by the user in the macro function \texttt{proto\_hls\_stream}, the corresponding set of HLS stream subroutines are generated for each of these types inside the module. For instance, the \emph{hls\_lowered\_write\_integer} subroutine at line 8 in Listing \ref{lst:module_instantiation} has been generated to accept an integer because that is the type provided to \texttt{proto\_hls\_stream}.


Similarly to the template instantiation mechanism in C/C++ HLS for typing streams, the macro at line 9 of Listing \ref{lst:code_example} is equivalent to \texttt{hls::stream<int> s}. This macro is expanded to \texttt{call set\_depth\_integer(s\%data\_integer, 0)} and has two purposes. Firstly, the type of the stream is added to a dictionary that keeps track of the stream types for each subroutine in the code. Secondly, the function call in the IR generated by Flang is replaced by the corresponding AMD Xilinx LLVM-IR primitive \texttt{@llvm.fpga.set.stream.depth}. This primitive is fundamental for AMD Xilinx's HLS back end to correctly handle streams, because otherwise it will attempt to implement these as Master AXI-4 interfaces.

\begin{lstfloat}
\begin{lstlisting}[language=Fortran, frame=lines, label=lst:module_instantiation, numbers=left, caption=Expansion of \texttt{proto\_hls\_stream} macro. Only the generated hls\_lowered\_write\_integer subroutine shown.]
module hls_stream
    type(HLSStream)
    end type

    type(HLSStream_integer)
        integer :: data_integer
    end type
    subroutine hls_lowered_write_integer(
        input_data, stream_data)
        integer, value :: input_data
        integer :: stream_data
    end subroutine
    ... 
end module
\end{lstlisting}
\end{lstfloat}

This approach of representing stream polymorphic types as a Fortran derived type, similar to a \emph{typedef} in C, with one field of the derived type being the type of the stream, is beneficial because we found that it ensures Flang generates similar LLVM-IR to Clang when accessing the member of a structure. This makes it more straightforward to locate the argument(s) of the AMD Xilinx's LLVM-IR primitive and/or the result of invoking such functions. 

\section{Evaluation} 
\label{sect:evaluation}

\begin{table*}[h]
  \begin{center}
  \caption{TeaLeaf mini-app kernels in HLS using C and Fortran (unoptimised) for the FPGA, also comparing against corresponding CPU performance.} 
  \label{tab:unoptimised}
  
  \begin{tabular}{|c|c|cc|cc|}
    \hline
    & & \multicolumn{2}{c|}{\textbf{Alveo U280 FPGA}} & \multicolumn{2}{c|}{\textbf{Xeon Platinum CPU}} \\
    \textbf{Benchmark} & \textbf{Problem size} & \makecell{\textbf{HLS C} \\ \textbf{Runtime(s)}} & \makecell{\textbf{HLS Fortran} \\ \textbf{Runtime(s)}} & \makecell{\textbf{C} \\ \textbf{Runtime(s)}} & \makecell{\textbf{Fortran} \\ \textbf{Runtime(s)}} \\
    \hline
    calc\_2norm & 480000 & 0.056 & 0.038 & 0.002 & 0.089 \\
    calc\_residual & 480000 & 0.196 & 0.215 & 0.008 & 0.087 \\
    cg\_calc\_p & 480000 & 0.315 & 0.319 & 0.002 & 0.003 \\
    cg\_calc\_ur & 480000 & 0.781 & 0.947 & 0.015 & 0.034 \\
    cg\_calc\_w & 480000 & 0.198 & 0.201 & 0.009 & 0.006 \\
    cg\_calc\_w\_norxy & 480000 & 0.317& 0.299 & 0.008 & 0.005 \\
    cg\_init & 480000 & 0.214 & 0.214 & 0.007 & 0.007 \\
    cheby\_init & 480000 & 0.850 & 0.845 & 0.025 & 0.093 \\
    cheby\_iterate & 480000 & 0.628 & 1.072 & 0.016 & 0.083 \\
    common\_init & 480000 & 0.711 & 0.631 & 0.025 & 0.093 \\
    field\_summary & 480000 & 0.105 & 0.054 & 0.008 & 0.095 \\
    finalise & 480000 & 0.003 & 0.055 & $\approx$0.000 & 0.107 \\
    generate\_chunk & 480000 & 0.369 & 0.898 & 0.008 & 0.096 \\
    initialise\_chunk & 480000 & 0.116 & 0.143 & $\approx$0.000 & 0.048 \\
    jacobi\_solve & 480000 & 0.192 & 0.230 & 0.018 & 0.001 \\
    ppcg\_calc\_rrn & 39677401 & 3.871 & 3.866 & 0.184 & 0.203 \\    
    ppcg\_calc\_zrnorm & 480000 & 0.059 & 0.059 & 0.002 & 0.086 \\
    ppcg\_init & 480000 & 0.343 & 0.509 & 0.009 & 0.109 \\
    ppcg\_init\_sd & 480000 & 0.227 & 0.188 & 0.005 & 0.069 \\
    ppcg\_inner & 480000 & 0.754 & 0.727 & $\approx$0.000 & 0.106 \\
    ppcg\_inner\_norxy & 480000 & 0.748 & 0.704 & $\approx$0.000 & 0.098 \\    
    ppcg\_pupdate & 480000 & 0.043 & 0.044 & 0.002 & 0.084 \\
    ppcg\_store\_r & 480000 & 0.043 & 0.044 & 0.002 & 0.086 \\    
    ppcg\_update\_z & 480000 & 0.043 & 0.044 & 0.002 & 0.099 \\
    set\_field & 480000 & 0.043 & 0.044 & 0.002 & 0.087 \\        
    tea\_block\_init & 480000 & 0.616 & 0.615 & 0.005 & 0.020 \\
    tea\_block\_solve & 480000 & 0.098 & 0.267 & $\approx$0.000 & 0.026 \\  
    tea\_diag\_init & 480000 & 0.083 & 0.083 & 0.002 & 0.019 \\
    tea\_diag\_solve & 480000 & 0.083 & 0.099 & 0.002 & 0.019 \\
    update\_halo & 480000 & 0.788 & 0.889 & 0.011 & 0.059 \\
    update\_halo\_cell & 480000 & 0.001 & 0.001 & $\approx$0.000 & 0.008 \\
    update\_in\_halo\_bt & 480000 & 0.016 & 0.016 & $\approx$0.000 & 0.067 \\
    update\_in\_halo\_cell\_bt & 480000 & 0.002 & 0.002 & $\approx$0.000 & 0.004 \\
    update\_in\_halo\_cell\_lr & 480000 & 0.049 & 0.080 & 0.001 & 0.006 \\
    update\_in\_halo\_lr & 480000 & 0.874 & 0.898 & 0.011 & 0.067 \\
  \hline
\end{tabular}
  \end{center}
\end{table*}

We evaluate the work presented in this paper with TeaLeaf \cite{mcintosh2017tealeaf}, an HPC mini-app that solves the linear heat conduction equation using 5 point stencils with implicit solvers. In total, this suite comprises 35 kernels which corresponds to 2587 lines of code ranging from Conjugate Gradient iteration to the Chebyshev method. Table \ref{tab:unoptimised} presents performance results for the baseline implementation of these kernels on an AMD Xilinx Alveo U280 FPGA and a single core of a Xeon Platinum CPU. It should be highlighted that the standard deviation across the 10 averaged runtimes is negligible, only affecting the 4th decimal figure of the results, and hence this has been omitted for brevity. It is not our intention in this paper to study the performance differences between architectures for these benchmarks and Table \ref{tab:unoptimised} reports a direct translation from the CPU to FPGA without any FPGA specific optimisations being undertaken because, in our first experiment, we are interested in the performance that our HLS Fortran tool can provide \emph{out of the box}. Indeed, the comparison between FPGA and CPU performance is provided to highlight where there are performance differences resulting from the activities of the front ends of Clang and Flang, for instance undertaking language specific code-level optimisations.

Several high level observations can be made from the runtime results reported in Table \ref{tab:unoptimised}. Firstly, it can be seen that there are performance differences between the HLS C and HLS Fortran kernels, but in the main these differences are fairly small. Furthermore, there is no clear pattern around whether C or Fortran performs best with HLS. For instance, the \emph{field\_summary} Fortran kernel outperforms C, whereas the opposite is true for the \emph{generate\_chunk} kernel. Table \ref{tab:overall_resource_usage} reports resource usage for these TeaLeaf kernels when using HLS C and HLS Fortran. The numbers in Table \ref{tab:overall_resource_usage} are expressed as the percentage of overall resources that are used on the Alveo U280, and it can be seen that whilst the BRAM resource usage between HLS C and HLS Fortran is fairly consistent, for the LUTs, FFs, and DSP slices it is more varied. In general terms, HLS Fortran tends to require more resources, especially LUTs, compared to HLS C but there are exceptions to this such as with the \emph{cg\_calc\_p} kernel.

\begin{table*}[h]
\begin{center}
\begin{tabular}{|c|cccc|cccc|}
\hline
&
\multicolumn{4}{c|}{\textbf{HLS C \% usage}}
& 
\multicolumn{4}{c|}{\textbf{HLS Fortran \% usage}}
\\

\cline{2-9}

\textbf{Kernel} & \textbf{BRAM} & \textbf{LUT} & \textbf{FF} & \textbf{DSP} 
& \textbf{BRAM} & \textbf{LUT} & \textbf{FF} & \textbf{DSP}\\
\hline
calc\_2norm & 0.10 & 0.32 & 0.14 & 0.19 & 0.10 & 0.43 & 0.16 & 0.16 \\
calc\_residual & 0.10 & 0.50 & 0.30 & 0.32 & 0.10 & 0.64 & 0.35 & 0.29 \\
cg\_calc\_p & 0.10 & 0.52 & 0.22 & 0.32 & 0.10 & 0.49 & 0.22 & 0.27 \\
cg\_calc\_ur & 0.10 & 2.24 & 1.20 & 1.30 & 0.10 & 2.88 & 1.23 & 1.00 \\
cg\_calc\_w & 0.10 & 0.69 & 0.36 & 0.42 & 0.10 & 0.80 & 0.39 & 0.32 \\
cg\_calc\_w\_norxy & 0.10 & 0.60 & 0.31 & 0.32 & 0.10 & 0.75 & 0.37 & 0.32 \\
cg\_init & 0.10 & 1.82 & 1.09 & 1.26 & 0.10 & 2.19 & 0.95 & 0.81 \\
cheby\_init & 0.10 & 2.15 & 1.31 & 1.43 & 0.10 & 2.50 & 1.19 & 0.98 \\
cheby\_iterate & 0.10 & 2.31 & 1.36 & 1.46 & 0.10 & 2.85 & 1.31 & 1.10 \\
common\_init & 0.10 & 2.29 & 1.42 & 1.13 & 0.10 & 2.82 & 1.53 & 0.79 \\
field\_summary & 0.10 & 0.44 & 0.24 & 0.22 & 0.10 & 0.56 & 0.26 & 0.27 \\
finalise & 0.10 & 0.22 & 0.11 & 0.10 & 0.10 & 0.30 & 0.13 & 0.03 \\
generate\_chunk & 0.10 & 1.05 & 0.54 & 0.32 & 0.10 & 1.63 & 0.74 & 0.70 \\
initialise\_chunk & 0.10 & 0.91 & 0.33 & 0.12 & 0.10 & 1.07 & 0.46 & 0.22 \\
jacobi\_solve & 0.10 & 0.63 & 0.37 & 0.35 & 0.10 & 0.85 & 0.44 & 0.32 \\
ppcg\_calc\_rrn & 0.10 & 0.37 & 0.17 & 0.19 & 0.10 & 0.45 & 0.20 & 0.23 \\
ppcg\_calc\_zrnorm & 0.10 & 0.35 & 0.15 & 0.19 & 0.10 & 0.46 & 0.18 & 0.16 \\
ppcg\_init & 0.10 & 1.95 & 1.15 & 1.33 & 0.10 & 2.39 & 1.04 & 0.84 \\
ppcg\_init\_sd & 0.10 & 0.54 & 0.26 & 0.22 & 0.10 & 0.68 & 0.30 & 0.20 \\
ppcg\_inner & 0.10 & 2.88 & 1.61 & 1.66 & 0.10 & 3.62 & 1.70 & 1.46 \\
ppcg\_inner\_norxy & 0.10 & 2.55 & 1.43 & 1.46 & 0.10 & 3.16 & 1.43 & 1.14 \\
ppcg\_pupdate & 0.10 & 0.23 & 0.09 & 0.07 & 0.10 & 0.31 & 0.11 & 0.03 \\
ppcg\_store\_r & 0.10 & 0.23 & 0.09 & 0.07 & 0.10 & 0.31 & 0.11 & 0.03 \\
ppcg\_update\_z & 0.10 & 0.23 & 0.09 & 0.07 & 0.10 & 0.31 & 0.11 & 0.03 \\
set\_field & 0.10 & 0.23 & 0.09 & 0.07 & 0.10 & 0.31 & 0.11 & 0.03 \\
tea\_block\_init & 0.10 & 0.78 & 0.35 & 0.48 & 0.10 & 0.89 & 0.35 & 0.35 \\
tea\_block\_solve & 0.10 & 1.23 & 0.82 & 0.93 & 0.10 & 1.30 & 0.65 & 0.61 \\
tea\_diag\_init & 0.10 & 0.26 & 0.38 & 0.07 & 0.10 & 0.36 & 0.42 & 0.07 \\
tea\_diag\_solve & 0.10 & 0.24 & 0.12 & 0.16 & 0.10 & 0.39 & 0.16 & 0.12 \\
update\_halo & 0.20 & 1.95 & 0.44 & 0.20 & 0.74 & 2.76 & 0.65 & 0.31 \\
update\_halo\_cell & 0.20 & 1.35 & 0.32 & 0.20 & 0.20 & 1.40 & 0.40 & 0.31 \\
update\_in\_halo\_bt & 0.74 & 1.48 & 0.34 & 0.13 & 0.74 & 1.67 & 0.42 & 0.17 \\
update\_in\_halo\_cell\_bt & 0.10 & 0.40 & 0.15 & 0.13 & 0.10 & 0.56 & 0.23 & 0.17 \\
update\_in\_halo\_cell\_lr & 0.10 & 0.39 & 0.18 & 0.13 & 0.10 & 0.50 & 0.22 & 0.13 \\
update\_in\_halo\_lr & 0.74 & 1.52 & 0.35 & 0.13 & 0.74 & 1.63 & 0.38 & 0.13 \\
\hline
\end{tabular}
\caption{Percentage of resource utilisation on the Alveo U280 FPGA for TeaLeaf mini-app kernels written in HLS using C or Fortran.}
\label{tab:overall_resource_usage}
\end{center}
\end{table*}

\begin{lstlisting}[language=c, frame=lines, label=lst:2d-c, caption=Access to a flattened 2D array with manually calculated offsets in C for \emph{cg\_calc\_w\_norxy} kernel.]
int x_dim = x_max - x_min;
int y_dim = y_max - y_min;
int w_x = x_max - x_min + 
    2 * halo_exchange_depth + 1;
for(int k=0; k <= y_dim; k++) {
    for(int j=0; j <= x_dim; j++) {
        w[(k + halo_exchange_depth) * w_x +
        (j + halo_exchange_depth)] = ...
    }
}    
\end{lstlisting}

Whilst the same algorithms from the TeaLeaf suite are being compiled using HLS C and HLS Fortran, the performance and resource usage differences between these flows reported in Tables \ref{tab:unoptimised} and \ref{tab:overall_resource_usage} demonstrate that there are implications of whether one uses C or Fortran to drive HLS. This is interesting because, based upon the fact that thus far we are using Von-Neumann based algorithms, then one would expect the unoptimised nature of these codes to dominate and performance to be very similar or the same between HLS C and HLS Fortran. Indeed our objective in this work has been for HLS Fortran to match the performance of HLS C. However, clearly the results are more nuanced and we focus on two kernels to explore these differences in more detail. Starting with the \emph{cg\_calc\_w\_norxy} kernel which undertakes Conjugate Gradient, as shown in Table \ref{tab:unoptimised}, this performs better on the FPGA in Fortran compared to C. The reason for this is because of the explicit calculation of offsets for accessing the flattened 2-dimensional array, as shown in Listing \ref{lst:2d-c}. Vitis imposes a limitation that external arrays have a maximum of one dimension, in C this must be handled manually by the programmer but with Fortran it is possible to use N-dimensional arrays. This is because dynamic arrays require pointer to pointers in C and are not supported by Vitis HLS \cite{xilinx2022vitis}. Arrays in Fortran are higher-level than in C, and-so during compilation Flang is able to calculate the offsets and flatten the dimensions of the arrays statically, rather than requiring this be undertaken at runtime, yielding an IR compatible with the aforementioned restriction. The programmer's Fortran code is illustrated in Listing \ref{lst:2d-fortran}, and this demonstrates an important advantage of using Fortran compared to C/C++ for programming FPGAs, where the additional expressiveness of Fortran, in this case via N-dimensional arrays, means that the compiler is able to undertake such offset calculations statically which are low level, time consuming, and can be error prone.

\begin{lstlisting}[language=fortran, frame=lines, label=lst:2d-fortran, caption=Access to a 2D array in Fortran for \emph{cg\_calc\_w\_norxy} kernel.]
REAL(KIND=8), 
DIMENSION(x_min-halo_exchange_depth:
          x_max+halo_exchange_depth,
          y_min-halo_exchange_depth:
          y_max+halo_exchange_depth) :: w

do k=y_min,y_max
    do j=x_min,x_max
        w(j,k) = ...
    end do
end do
\end{lstlisting}

The \emph{ppcg\_calc\_rrn} kernel, which undertakes a reduction and whose runtime is reported in Table \ref{tab:unoptimised}, is also slightly faster using our Fortran-based approach compared to C on the FPGA. It is instructive to compare the runtime between C and Fortran on the FPGA against that of the CPU to understand whether performance differences come from these respective front ends. For instance, it can be seen that for the \emph{cg\_calc\_w\_norxy} kernel the Fortran runtime is lower than the C runtime for both the FPGA and CPU, whereas this is not the case for the \emph{ppcg\_calc\_rrn} kernel where Fortran is slightly faster than C on the FPGA but slower on the CPU. It can be seen that there is no clear pattern, and just because one language is faster than the other on the FPGA then it does not mean that the same will be true on the CPU, or vice versa. Consequently, it can be deduced that it is the combination of both the front end and back end which is determining runtime performance and differences between languages are at the individual kernel level.

Results reported in Table \ref{tab:unoptimised} are unoptimised for the FPGA, and whilst this was an important baseline to understand the performance differences delivered for such codes, it is well known that Von-Neumann based codes must be transformed to their dataflow counterparts for best performance on an FPGA \cite{brown2021porting}. A key question was therefore whether support in our tool was rich enough to enable these same optimisations to be undertaken at the code level in HLS via Fortran, as is currently possible in C or C++. We focus on optimising the \emph{cg\_calc\_w\_norxy} and \emph{ppcg\_calc\_rrn} kernels, with the results of optimising these kernels reported in Table \ref{tab:optimised}. We optimised the \emph{cg\_calc\_w\_norxy} Conjugate Gradient stencil kernel using shift registers. This technique replaces the strided memory accesses by sequential accesses to a shift register, where every element can be accessed in one cycle due to the initial prefetching of part of the array and a sliding window, where the access to new elements from memory is pipelined via the shift register operation \cite{brown2021accelerating}. It can be seen that this optimisation reduces the runtime of both the C and Fortran versions compared to their unoptimised counterparts.

\begin{table}[h]
  \begin{center}
  \caption{Optimised kernels in HLS C and HLS Fortran for the FPGA, and also a synthetic streaming kernel. Problem size of 480000 for \emph{cg\_calc\_w\_norxy}, 39677401 for \emph{ppcg\_calc\_rrn}, and 32 million elements for Synthetic Streaming.}
  \label{tab:optimised}
  
  \begin{tabular}{|c|cc|}
    \hline
    & \multicolumn{2}{c|}{\textbf{Alveo U280 FPGA}} \\
    \textbf{Benchmark} & \makecell{\textbf{HLS C} \\ \textbf{Runtime(s)}} & \makecell{\textbf{HLS Fortran} \\ \textbf{Runtime(s)}} \\
    \hline
    \makecell{cg\_calc\_w\_norxy}  & 0.200 & 0.222\\
    \makecell{ppcg\_calc\_rrn (partial sums)} & 0.630 & 0.631\\    
    \makecell{ppcg\_calc\_rrn (dataflow)} & 0.820 & 0.535\\
    \makecell{Synthetic Streaming} & 0.131 & 0.132\\
  \hline
\end{tabular}
  \end{center}
\end{table}

The bottleneck with the \emph{ppcg\_calc\_rrn} reduction kernel is in the floating point adder, which requires 7 cycles to complete a sum, therefore leading to an initiation interval of 7 because results from consecutive sum operations are required. We ameliorate this with partial sums, where Vitis pipelines the adder to undertake seven independent sum operations every 7 cycles instead \cite{hrica2012floating}, and then at the end adds these seven partial sums together to calculate the overall result.

There are two versions of the \emph{ppcg\_calc\_rrn} kernel reported in Table \ref{tab:optimised}, \emph{partial sums} uses the partial sum technique only and it can be seen that both Fortran and C deliver very similar performance on the FPGA, which is significantly faster than the unoptimised version in Table \ref{tab:unoptimised}. The second version of the \emph{ppcg\_calc\_rrn} kernel, \emph{dataflow}, uses the \emph{dataflow} pragma and HLS streams to pipeline both the compute and interaction with external memory to load and store data. 

This second optimised version of the \emph{ppcg\_calc\_rrn} kernel, using dataflow regions and HLS streams, yielded surprising results, as it can be seen from Table \ref{tab:optimised} that the Fortran version is 53.27\% faster than the C HLS code. Based upon the report generated by Vitis HLS, we observed that Vitis is unable to pipeline the reduction loop when using C, whereas it is able to do so in Fortran. Because of the failure of Vitis to pipeline the loop in C, the performance of the dataflow optimisation does not improve with respect to the partial sums. Whilst our kernels all ran at 300Mhz on the FPGA, it should also be noted that the maximum operating frequency reported by Vitis HLS is also different, up to 371.47 MHz possible with Fortran and 357.68 MHz for C. It is interesting to observe that small differences in the LLVM-IR resulting from different language front ends, can have a significant impact. 

Finally, we developed a synthetic example, Synthetic Streaming in Table \ref{tab:optimised},  which sums two vectors from external memory as data is streamed in. This kernel relies on three dataflow regions connected via HLS streams; loading data from external memory, undertaking the element sum, and writing results back to external memory.  Compared to the \emph{ppcg\_calc\_rrn} kernel, these loops are much simpler and-so can be pipelined by Vitis when using either the C and Fortran flows, and both languages deliver similar performance.

\begin{table}[h]
  \begin{center}
  \caption{Percentage total resource usage in the optimised kernels and synthetic streaming in HLS C and HLS Fortran. Percentage overhead of Fortran also reported in brackets, in green the cases where Fortran leads to less resource utilisation.}
  \label{tab:optimised_resources}
  
  \begin{tabular}{|c|cccc|cccc|}
    \hline
    & \multicolumn{4}{c|}{\textbf{HLS kernel usage C/Fortran (\% total resources)}} \\
    \textbf{Benchmark} & LUTs & FFs & BRAM & DSP \\
    \hline
    \makecell{cg\_calc\_w\\\_norxy}  & 6.13 / 5.76 & 6.32 / 5.13 & 3.17 / 3.17 & 0.74 / 0.74\\
    \makecell{} & \textcolor{green}{(-6.04\%)} & \textcolor{green}{(-18.83\%)} & (0.00\%) & (0.00\%) \\
    \makecell{ppcg\_calc\_rrn \\(partial sums)} & 0.65 / 0.60 & 0.34 / 0.30 & 0.40 / 0.40 & 0.19 / 0.22\\
    \makecell{} & \textcolor{green}{(-7.69\%)} & \textcolor{green}{(-11.76\%)} & (0.00\%) & (15.79\%) \\
    \makecell{ppcg\_calc\_rrn \\(dataflow)} & 0.64 / 0.78 & 0.34 / 0.40 & 0.40 / 0.40 & 0.19 / 0.39 \\
    \makecell{} & (21.88\%) & (17.65\%) & (0.00\%) & (105.26\%) \\
    \makecell{Synthetic \\ Streaming} & 0.29 / 0.22  & 0.10 / 0.08 & 0.15 / 0.10 & 0.00 / 0.00\\
    \makecell{} & \textcolor{green}{(-24.13\%)} & \textcolor{green}{(-20.00\%)} & \textcolor{green}{(-33.33\%)} & - \\
  \hline
\end{tabular}
  \end{center}
\end{table}

Another important consideration is the resource utilisation that results from these tools when undertaking code optimisation, and whether one approach is more frugal than another. Table \ref{tab:optimised_resources} reports resource utilisation for the four optimised benchmarks that were explored in Table \ref{tab:optimised} and these can be compared against the resource utilisation for unoptimised kernels reported in Table \ref{tab:overall_resource_usage}. For each benchmark and resource, the percentage of overall resource usage is reported for the C and then the Fortran code, with these separated by a slash. Underneath, in brackets, the percentage overhead of Fortran compared to C is reported, where green is used to highlight situations where Fortran resulted in reduced resource usage.

It can be seen in Table \ref{tab:optimised_resources} that for both \emph{cg\_calc\_w\_norxy} and \emph{ppcg\_calc\_rrn} with partial sums kernels, Fortran requires fewer LUTs and FFs than C, with BRAM and DSP being comparable. With the synthetic streaming benchmark, where performance was comparable between the C and Fortran versions, it can be seen that Fortran also uses fewer resources.

Fortran requires resources than C for the \emph{ppcg\_calc\_rrn} with dataflow kernel, especially for DSP slices, however the hardware that is generated is fundamentally different between these versions due to the lack of pipelining on the reduction loop in C which led to lower performance. Therefore, overall, these utilisation results demonstrate that Fortran is competitive against C for resource utilisation when optimising the kernel and, in some cases, such as synthetic streaming, can result in fewer resources with no loss in performance.

\section{Conclusions and further work}
\label{sec:conclusions}
The work described in this paper enables programmers to leverage Fortran when programming AMD Xilinx FPGAs, which is especially important for HPC workloads as many of these are written in Fortran. Whilst Fortran programmers still need to transform their Von-Neumann algorithms to a dataflow representation via algorithmic restructuring and apply appropriate HLS pragmas, crucially our tool avoids the time consuming and error prone preliminary step of converting code into C/C++. This, combined with the fact that code remains in one single language which HPC developers are familiar with, is beneficial in encouraging HPC developers to explore accelerating their Fortran codes using FPGAs. The tool that we have described connects LLVM's Flang with the AMD Xilinx HLS back end, without any modification required to either Flang or the back end. As a result, the programmer is, apart from our conversion utility, using standard tools that are maintained by the wider community. There is no divergence from LLVM Flang or the AMD Xilinx HLS back end, and programmers benefit from enhancements made both in Flang by the community and in the HLS back end by AMD Xilinx. Furthermore, wider features of the Vitis ecosystem, such as profiling, are also made available to the Fortran programmer.

We have compared performance of our approach against the existing C/C++ Vitis HLS tooling. It was observed that, in the main, performance between these tools is fairly comparable, although there are some differences for instance with Fortran where the programmer can leverage higher level features, such as N-dimensional arrays, that improves expressiveness and enables the compiler to undertake additional optimisations. The major benefit of our approach is that we unlock the ability for large HPC applications that are written in Fortran to be run on FPGAs, and as we demonstrated these pragmas currently supported by our tool enables these codes to be optimised for FPGAs, with additional pragmas and constructs being trivial to add to the flow if required.

An objective has been for our HLS Fortran flow to match the performance of HLS C so that HPC programmers are able to accelerate their Fortran codes on FPGAs without any reduction in performance compared to leveraging C/C++. However, it can be observed that some HLS Fortran kernels execute slightly slower than their HLS C versions and consequently, for further work, optimisation at the LLVM-IR level would be beneficial. For instance, the \emph{cg\_calc\_w\_norxy} Fortran kernel performs slightly worse than its C counterpart and we believe that this is because of the use of long (64-bit) integers for array indicies in the IR generated by Flang. Currently these require integer conversion from the programmer's 32-bit indicies to 64-bit in the IR, and this could be optimised by our tool to avoid such conversions. It would also be interesting to apply our code to programming the AI engines that are present on the Versal, potentially providing a unified programming interface between the PL and AI engines. Because AMD Xilinx have already developed MLIR dialects for their AI engine compiler, and internally Flang uses MLIR via the FIR dialect, it should be possible to connect these tools and leverage our HLS Fortran flow. Furthermore, it would be interesting to explore generalising our tool, for instance to support other programming languages, such as Python, Rust or Julia, to be integrated with the AMD Xilinx back end.

We intend to make our tool open source and available at \emph{https://fpga.epcc.ed.ac.uk/community/fortran.html} .

\section*{Acknowledgment}

The authors would like to thank HPE who funded this work via the EMEA Research Lab internship programme, the ExCALIBUR xDSL project, and the ExCALIBUR H\&ES FPGA testbed and AMD Xilinx HACC program for access to compute resource used in this work. 

\bibliographystyle{IEEEtran}
\bibliography{references.bib}
\end{document}